\documentclass[reprint,superscriptaddress,amsmath,amssymb,aps,prc,twocolumn,nofootinbib]{revtex4-2}

\usepackage{graphicx}
\usepackage{xcolor}
\usepackage[english]{babel}
\usepackage{bm}
\usepackage[T1]{fontenc}
\usepackage{multirow}

\usepackage{lineno,hyperref}

\usepackage[output-decimal-marker={.},exponent-product=\cdot]{siunitx}

\begin{document}

\title{Mass measurements in the $^{132}$Sn region with the JYFLTRAP double Penning trap mass spectrometer}

\author{O.~Beliuskina}
\email{olga.o.beliuskina@jyu.fi}
\affiliation{University of Jyvaskyla, Department of Physics, Accelerator Laboratory, P.O. Box 35(YFL) FI-40014 University of Jyvaskyla, Finland}
\author{D.A.~Nesterenko}%
\affiliation{University of Jyvaskyla, Department of Physics, Accelerator Laboratory, P.O. Box 35(YFL) FI-40014 University of Jyvaskyla, Finland}
\author{A.~Jaries}
\affiliation{University of Jyvaskyla, Department of Physics, Accelerator Laboratory, P.O. Box 35(YFL) FI-40014 University of Jyvaskyla, Finland}
\author{M.~Stryjczyk}
\email{marek.m.stryjczyk@jyu.fi}
\affiliation{University of Jyvaskyla, Department of Physics, Accelerator Laboratory, P.O. Box 35(YFL) FI-40014 University of Jyvaskyla, Finland}
\author{A.~Kankainen}
\email{anu.kankainen@jyu.fi}
\affiliation{University of Jyvaskyla, Department of Physics, Accelerator Laboratory, P.O. Box 35(YFL) FI-40014 University of Jyvaskyla, Finland}
\author{L.~Canete}
\altaffiliation[Present address: ]{Northeastern University London, Devon House, 58 St Katharine's Way, E1W 1LP, London, United Kingdom}
\affiliation{University of Jyvaskyla, Department of Physics, Accelerator Laboratory, P.O. Box 35(YFL) FI-40014 University of Jyvaskyla, Finland}
\author{R.P.~de~Groote}
\altaffiliation[Present address: ]{KU Leuven, Instituut voor Kern- en Stralingsfysica, B-3001 Leuven, Belgium}
\affiliation{University of Jyvaskyla, Department of Physics, Accelerator Laboratory, P.O. Box 35(YFL) FI-40014 University of Jyvaskyla, Finland}
\author{C.~Delafosse}
\altaffiliation[Present address: ]{Universit\'e Paris Saclay, CNRS/IN2P3, IJCLab, 91405 Orsay, France}
\affiliation{University of Jyvaskyla, Department of Physics, Accelerator Laboratory, P.O. Box 35(YFL) FI-40014 University of Jyvaskyla, Finland}
\author{T.~Eronen}
\affiliation{University of Jyvaskyla, Department of Physics, Accelerator Laboratory, P.O. Box 35(YFL) FI-40014 University of Jyvaskyla, Finland}
\author{Z.~Ge}
\affiliation{University of Jyvaskyla, Department of Physics, Accelerator Laboratory, P.O. Box 35(YFL) FI-40014 University of Jyvaskyla, Finland}
\author{S.~Geldhof}
\altaffiliation[Present address: ]{GANIL, CEA/DRF-CNRS/IN2P3, B.P. 55027, 14076 Caen, France}
\affiliation{University of Jyvaskyla, Department of Physics, Accelerator Laboratory, P.O. Box 35(YFL) FI-40014 University of Jyvaskyla, Finland}
\author{W.~Gins}
\affiliation{University of Jyvaskyla, Department of Physics, Accelerator Laboratory, P.O. Box 35(YFL) FI-40014 University of Jyvaskyla, Finland}
\author{M.~Hukkanen}
\affiliation{University of Jyvaskyla, Department of Physics, Accelerator Laboratory, P.O. Box 35(YFL) FI-40014 University of Jyvaskyla, Finland}
\affiliation{Universit\'e de Bordeaux, CNRS/IN2P3, LP2I Bordeaux, UMR 5797, F-33170 Gradignan, France}
\author{A.~Jokinen}
\affiliation{University of Jyvaskyla, Department of Physics, Accelerator Laboratory, P.O. Box 35(YFL) FI-40014 University of Jyvaskyla, Finland}
\author{I.D.~Moore}
\affiliation{University of Jyvaskyla, Department of Physics, Accelerator Laboratory, P.O. Box 35(YFL) FI-40014 University of Jyvaskyla, Finland}
\author{M.~Mougeot}
\affiliation{University of Jyvaskyla, Department of Physics, Accelerator Laboratory, P.O. Box 35(YFL) FI-40014 University of Jyvaskyla, Finland}
\author{S.~Nikas}%
\affiliation{University of Jyvaskyla, Department of Physics, Accelerator Laboratory, P.O. Box 35(YFL) FI-40014 University of Jyvaskyla, Finland}
\author{H.~Penttil\"a}
\affiliation{University of Jyvaskyla, Department of Physics, Accelerator Laboratory, P.O. Box 35(YFL) FI-40014 University of Jyvaskyla, Finland}
\author{I.~Pohjalainen}
\affiliation{University of Jyvaskyla, Department of Physics, Accelerator Laboratory, P.O. Box 35(YFL) FI-40014 University of Jyvaskyla, Finland}
\author{A.~Raggio}
\affiliation{University of Jyvaskyla, Department of Physics, Accelerator Laboratory, P.O. Box 35(YFL) FI-40014 University of Jyvaskyla, Finland}
\author{M.~Reponen}
\affiliation{University of Jyvaskyla, Department of Physics, Accelerator Laboratory, P.O. Box 35(YFL) FI-40014 University of Jyvaskyla, Finland}
\author{S.~Rinta-Antila}
\affiliation{University of Jyvaskyla, Department of Physics, Accelerator Laboratory, P.O. Box 35(YFL) FI-40014 University of Jyvaskyla, Finland}
\author{A.~de~Roubin}
\altaffiliation[Present address: ]{LPC Caen, Normandie Univ., 14000 Caen, France}
\affiliation{University of Jyvaskyla, Department of Physics, Accelerator Laboratory, P.O. Box 35(YFL) FI-40014 University of Jyvaskyla, Finland}
\author{J. Ruotsalainen}
\affiliation{University of Jyvaskyla, Department of Physics, Accelerator Laboratory, P.O. Box 35(YFL) FI-40014 University of Jyvaskyla, Finland}
\author{M.~Vilen}
\affiliation{University of Jyvaskyla, Department of Physics, Accelerator Laboratory, P.O. Box 35(YFL) FI-40014 University of Jyvaskyla, Finland}
\author{V.~Virtanen}
\affiliation{University of Jyvaskyla, Department of Physics, Accelerator Laboratory, P.O. Box 35(YFL) FI-40014 University of Jyvaskyla, Finland}
\author{A.~Zadvornaya}
\affiliation{University of Jyvaskyla, Department of Physics, Accelerator Laboratory, P.O. Box 35(YFL) FI-40014 University of Jyvaskyla, Finland}

\date{\today}

\begin{abstract}
We report on new precision mass measurements of neutron-rich $^{137}$Sb and $^{136-142}$I isotopes from the JYFLTRAP double Penning trap mass spectrometer. We confirm the value from the previous Penning-trap measurement of $^{137}$Sb at the Canadian Penning Trap and therefore rule out the conflicting result from the Experimental Storage Ring. The ground state and isomer in $^{136}$I were resolved and measured directly for the first time. The isomer excitation energy, $E_x = 215.1(43)$~keV, agrees with the literature but is three times more precise. The measurements have improved the precision of the mass values and confirmed previous results in the majority of cases. However, for $^{138,140}$I the results differ by 17(6)~keV and 23(12)~keV, respectively. This could be explained by an unresolved contamination or different ratio of unresolved isomeric states in the case of $^{140}$I.
\end{abstract}

\maketitle

\section{Introduction}

\begin{figure}
\centering
\includegraphics[width=\columnwidth]{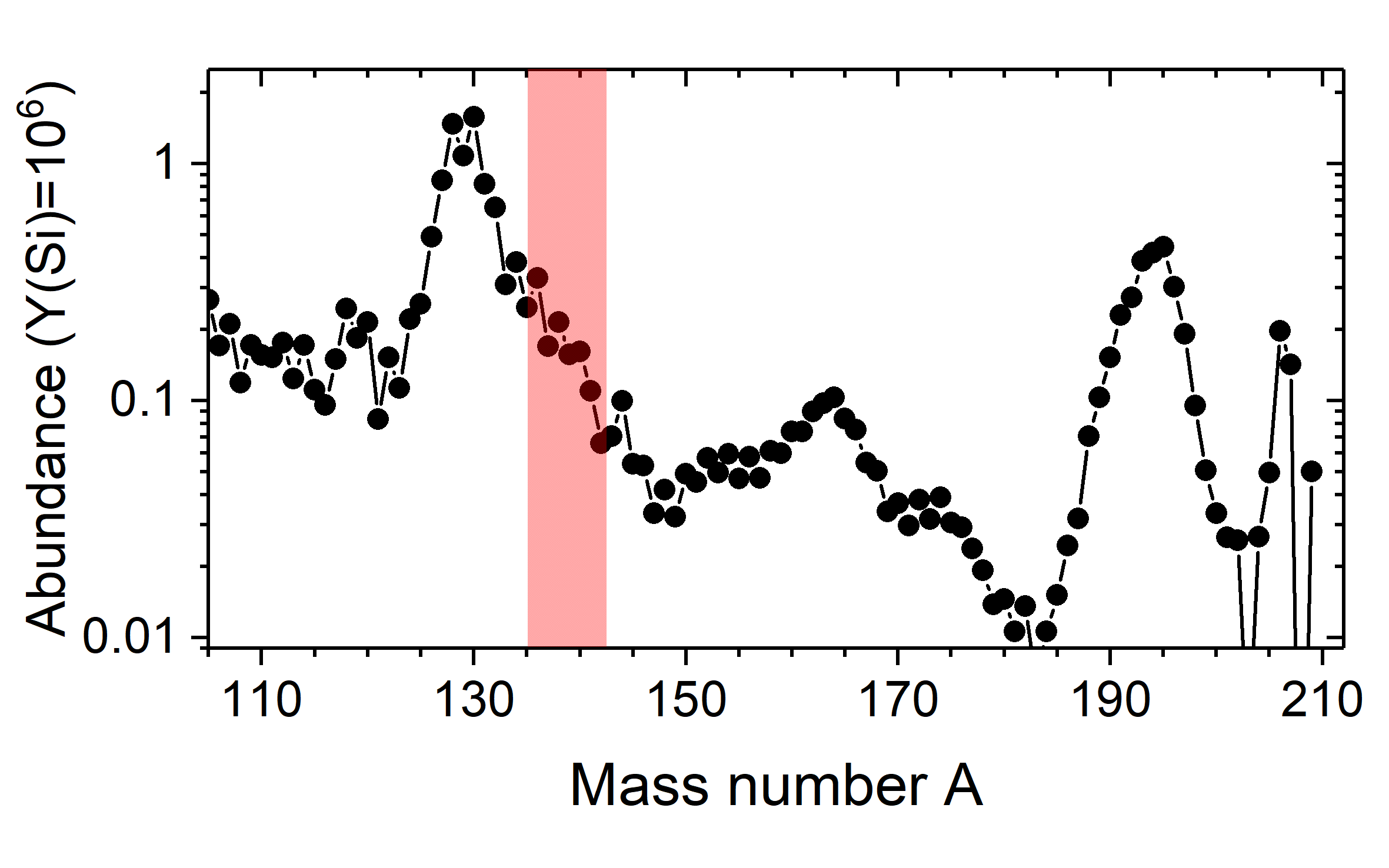}
\caption{\label{fig:abundances}Solar r-process abundances from Ref.~\cite{Goriely1999}. The region of interest (highlighted in red) is located in the abundance peak region around $A\approx130$.}
\end{figure}

The rapid neutron capture process (r process) \cite{Burbidge1957} produces around half of the heavy-element abundances above iron (see e.g. the recent review \cite{Cowan2021} and references therein). One of the regions of interest and of high abundance is located north-east from the doubly magic $^{132}$Sn, around $A=136-142$, next to the second r-process peak at $A\approx130$ (see Fig.~\ref{fig:abundances}). The astrophysical calculations, which attempt to reproduce the observed r-process abundances, depend on input parameters from nuclear physics experiments on radioactive nuclei, such as atomic masses \cite{Mumpower2016}. In the discussed region, it has been shown that the mass uncertainties as low as 20 keV can affect the calculated neutron-capture reaction rates \cite{Nikas2022}. Yet, despite their importance, many of the known masses of iodine isotopes are based only on a single measurement \cite{Huang2021} performed at the Canadian Penning Trap (CPT) \cite{VanSchelt2012,VanSchelt2013,Orford2020}. 

While Penning trap mass spectrometry offers the most precise method to determine masses of radioactive isotopes, disagreements between different Penning-trap measurements have been found in the past due to unaccounted systematic effects, for example in $^{91}$Sr \cite{Raimbault-Hartmann2002,Jaries2024}, unidentified beam contaminants, as in $^{140}$Te \cite{Hakala2012,VanSchelt2013,Orford2018}, or presence of unknown isomeric states, e.g. in $^{162}$Eu \cite{Vilen2018,Hartley2018,Vilen2020a} and $^{163}$Gd \cite{VanSchelt2012,Vilen2018,Vilen2020a,Vilen2020err,Orford2020}. On top of that, discrepancies between results from different experimental techniques have also been reported, a striking example of which is $^{137}$Sb. Its mass was measured using Isochronous Mass Spectrometry at the Experimental Storage Ring (ESR) at GSI \cite{Knobel2016}. However, the reported value deviated by 549(181)~keV with respect to the earlier CPT result \cite{VanSchelt2013}. Since $^{137}$Sb has been highlighted as one of the key r-process nuclei for which the neutron separation energy has to be accurately determined \cite{Brett2012}, a new high-precision mass measurement is relevant. Thus, there is a clear need for cross-checking known values via independent studies at different facilities. 

Some of the iodine isotopes in the region of $^{132}$Sn have isomeric states that can impact the mass measurements if not properly resolved and identified. For example, $^{136}$I has a long-living isomeric state at 206(15)~keV \cite{NUBASE20} but its ground state is determined only from decay studies \cite{Fogelberg2007,AME2020} that are known to be prone to systematic shifts \cite{Goncharov2011,Nesterenko2019,Ramalho2022,Gamage2022a,Hager2007,Hukkanen2023,Audi2012}. In the case of $^{140}$I, a recent $\beta$-decay study performed at RIKEN \cite{Yagi2022} revealed the presence of three long-lived states. While the time-of-flight ion cyclotron resonance (TOF-ICR) technique \cite{Konig1995,Graff1980} previously used at CPT did not provide high enough resolving power in the measurement of iodine isotopes to resolve all isomeric states, the phase-imaging ion cyclotron resonance (PI-ICR) technique \cite{Eliseev2014,Nesterenko2018} has been proven to be able to resolve low-lying states only a few tens of keV apart \cite{Orford2018,Nesterenko2020,Nesterenko2022,Hukkanen2023,Ruotsalainen2023,Jaries2023}.

In this work, we report on mass measurements of neutron-rich $^{136-142}$I and $^{137}$Sb isotopes. These measurements were performed at the JYFLTRAP double Penning trap \cite{Eronen2012} using both the TOF-ICR and PI-ICR technique in order to check the consistency with the previous measurements and to provide more accurate mass values. 

\section{Experimental method}

The mass measurements were performed at the Ion Guide Isotope Separator On-Line facility (IGISOL) at the University of Jyv\"askyl\"a \cite{Moore2013} using the JYFLTRAP double Penning trap mass spectrometer~\cite{Eronen2012} during three different experiments. The isotopes of $^{136,138}$I and $^{137}$Sb were studied during Run~I, $^{136-142}$I in Run~II and $^{140}$I in Run~III. In all three experiments radioactive species were produced in proton-induced fission of a thin (15~mg/cm$^{2}$) $^{nat}$U target utilizing 30- (Run I) or 25-MeV (Run II and III) protons delivered by the K-130 cyclotron. The reaction products were thermalized in helium buffer gas, typically at 250-300~mbar, extracted using a radio-frequency sextupole ion guide \cite{Karvonen2008} and, subsequently, accelerated to 30 keV. The singly-charged ions of interest were first selected with respect to the mass-over-charge ratio using a dipole magnet and then the continuous beam was cooled and bunched by a gas-filled radio-frequency quadrupole cooler and buncher \cite{Nieminen2001}. From there, the ion bunches were injected into the JYFLTRAP double Penning trap.

JYFLTRAP is comprised of two cylindrical Penning traps housed in a 7-T superconducting magnet \cite{Gabrielse1984,Eronen2012,Kolhinen2004}. In the purification trap, ions were cooled, centered and purified via a mass-selective buffer gas cooling technique \cite{Savard1991}. In case of $^{136}$I$^{m}$ measured in Run I, the Ramsey cleaning method \cite{Eronen2008} with an excitation pattern (on-off-on) 5-380-5 ms was additionally used to separate the isomer from the ground state. The centered ions were extracted through the small aperture from the purification trap to the precision trap, where the ion cyclotron frequencies are measured. For the measurements reported in this work two methods were used, TOF-ICR \cite{Konig1995,Graff1980} and PI-ICR \cite{Eliseev2014,Nesterenko2018}.

The ion mass is related to the measured cyclotron frequency $\nu_{c}$:
\begin{equation}
\nu_{c} = \frac{1}{2\pi}\frac{q}{m}B \mathrm{,}
\end{equation}
where $B$ is the magnetic field strength and $q/m$ is the charge-to-mass ratio of the measured ion. In order to determine the magnetic field strength precisely, the cyclotron frequency of a well-known reference ion was measured. During Run I, two isotopes of xenon, $^{130}$Xe (literature mass-excess value ${\Delta_{lit.} = -89880.474(9)}$~keV~\cite{AME2020}) and $^{136}$Xe (${\Delta_{lit.} = -86429.170(7)}$~keV~\cite{AME2020}) were used as references while during Run II and III, stable $^{133}$Cs (${\Delta_{lit.} = -88070.943(8)}$~keV~\cite{AME2020}) delivered from the IGISOL offline surface ion source \cite{Vilen2020} was utilized. The measurements of the ion of interest and the reference ions were alternated to account for the temporal magnetic field fluctuations. 

The atomic mass of the ion of interest $M$ can be extracted from the ratio of cyclotron frequencies ${r = \nu_{c,ref}/\nu_{c}}$ as follows:
\begin{equation}
M = r(M_{ref} - m_e) + m_e\mathrm{,}
\end{equation}
where $m_e$ and $M_{ref}$ are the mass of an electron and the atomic mass of the reference ion, respectively. For species measured against the isobaric reference the energy difference $Q$ between them was calculated as:
\begin{equation}
Q = (r-1)[M_{ref} - m_e]c^2 \mathrm{,}
\end{equation} 
where $c$ is the speed of light in vacuum.

In the TOF-ICR technique \cite{Konig1995,Graff1980}, the ion’s initial magnetron motion is converted into cyclotron motion by applying a quadrupole excitation pulse with a fixed duration and amplitude. This frequency is scanned around the expected cyclotron frequency of the ion while recording the ions’ time-of-flight. It results in a variable degree of conversion of the slow magnetron motion into the fast reduced cyclotron motion depending on the excitation frequency and, consequently, an increase in the associated radial energy. This increase is observed as a shorter time-of-flight of the ions from the precision trap onto a microchannel plate (MCP) detector located after the trap.

\begin{figure}
\centering
\includegraphics[width=\columnwidth]{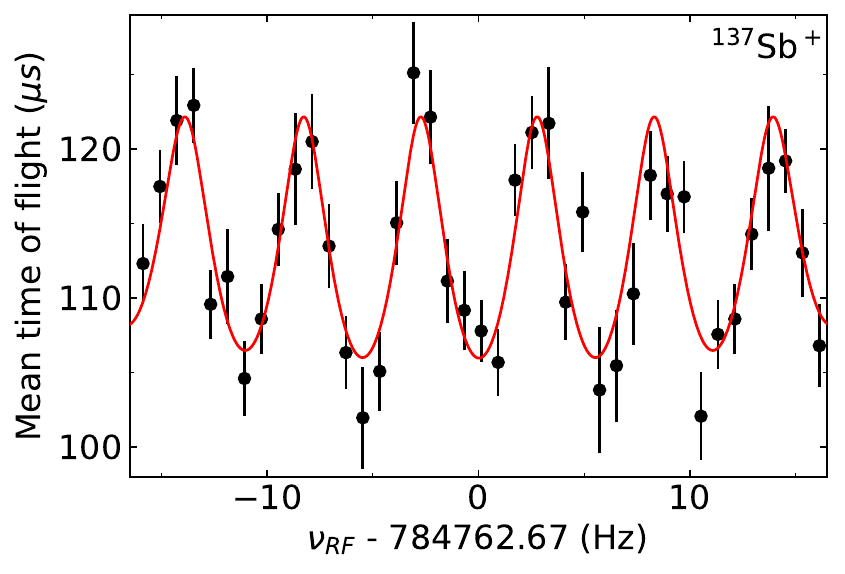}
\caption{\label{fig:Ramsey}TOF-ICR spectrum for $^{137}$Sb$^+$ using a 25-150-25~ms (on-off-on) Ramsey excitation pattern. The black squares with error bars are the mean time-of-flight values for each scanned frequency and the solid red line is a fit of the theoretical curve \cite{Kretzschmar2007} to the data points.}
\end{figure}

\begin{figure}
\centering
\includegraphics[width=\columnwidth]{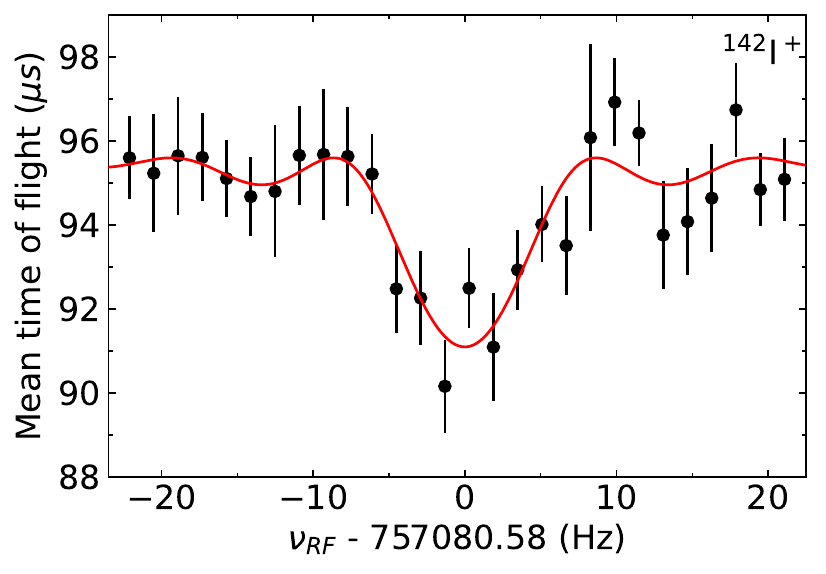}
\caption{\label{fig:142I}Example of a TOF-ICR spectrum for $^{142}$I$^+$ using a 100~ms quadrupolar one-excitation pulse. The black squares with error bars are the mean time-of-flight values for each scanned frequency and the solid red line is a fit of the theoretical curve to the data points.}
\end{figure}

With the exception of $^{142}$I, Ramsey's method of time-separated oscillatory fields \cite{Kretzschmar2007,George2007} was applied for all cases measured with TOF-ICR. The excitation pattern comprised of two short pulses separated by a period without excitation. Three on-off-on patterns were used, 25-750-25~ms for $^{136}$I$^{m}$ and $^{137}$I, 25-350-25~ms for $^{137-140}$I and 25-150-25~ms for $^{137}$Sb and $^{141}$I, see Fig.~\ref{fig:Ramsey}. The choice of employed excitation scheme depends on the half-life, production rates and possible isomeric contaminants. In the case of $^{142}$I, a conventional quadrupolar one-excitation pulse of 100~ms was used, see Fig.~\ref{fig:142I}. 

\begin{figure}
\centering
\includegraphics[width=\columnwidth]{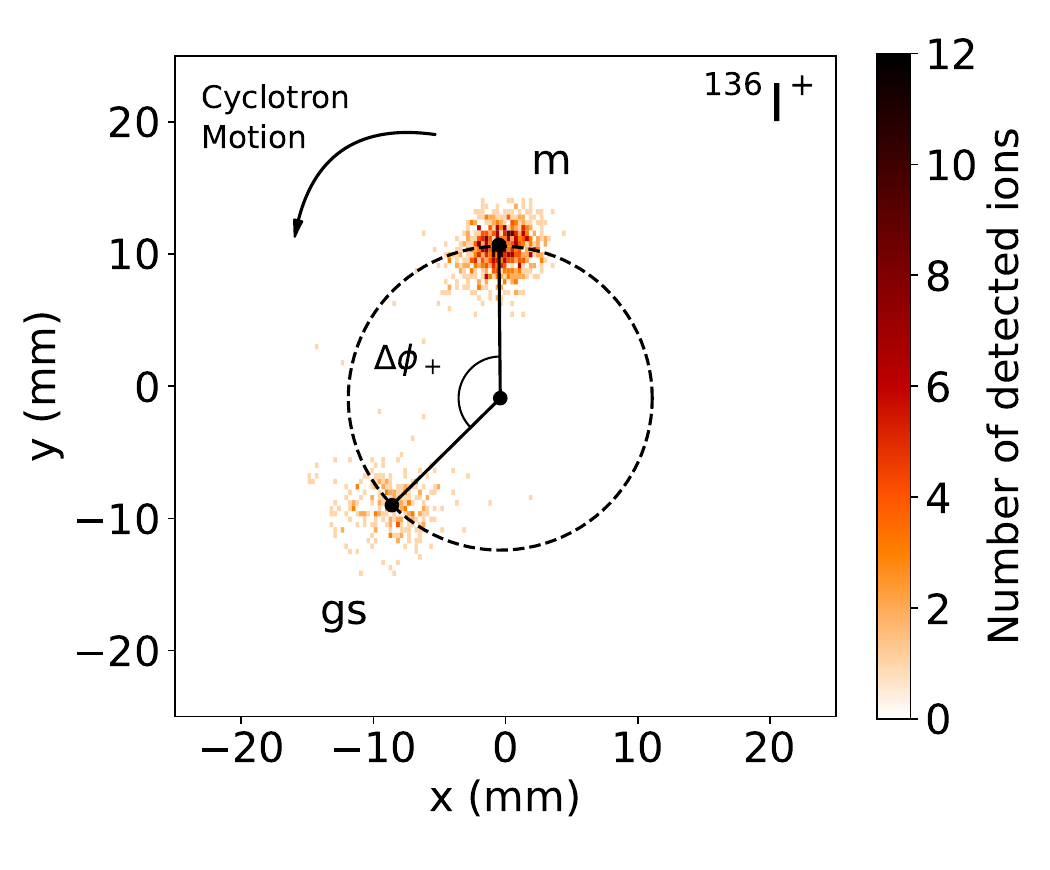}
\caption{\label{fig:136I}Projection of the cyclotron motion of $^{136}$I ions onto the position-sensitive detector obtained with the PI-ICR technique using a phase accumulation time $t_{acc} = 300$~ms. The average excitation radius is indicated with the dashed circle while the position of the center spot (not shown here) and the cyclotron spots are marked with the $\bullet$ symbol. The angle difference $\Delta\phi_+$ leads to an excitation energy ${E_x = 215.1(43)}$~keV.}
\end{figure}

In the PI-ICR method \cite{Eliseev2014, Nesterenko2018}, the cyclotron frequency relies on detecting the projections of the ion’s radial motion phases after a certain phase accumulation time $t_{acc}$, projected onto a position-sensitive MCP detector (2D~MCP). Final phases of both magnetron ($\phi_-$) and cyclotron ($\phi_+$) are needed, requiring two separate measurement patterns. Both patterns start with an ion excitation with a short dipolar pulse to establish cyclotron motion. During the measurement of the cyclotron phase, the ion is allowed to accumulate a phase on the modified cyclotron motion for a given time $t_{acc}$ and then it is converted with a $\pi$ pulse to a magnetron motion. In the measurement of the magnetron phase, the initial cyclotron motion is immediately converted to a magnetron motion and allowed to accumulate the $\phi_-$ phase for time $t_{acc}$. The ion cyclotron frequency is related to the measured phase difference $\phi_c = \phi_+ - \phi_-$ and it follows the equation: 
\begin{equation}
\nu_c = (\phi_c + 2\pi n) / 2 \pi t_{acc},
\end{equation}
where $n$ is the number of the full revolutions of the ion in the precision trap. In this work, the PI-ICR technique was used during Run II to measure $^{136}$I ($t_{acc} = 300$~ms) and during Run III for $^{140}$I ($t_{acc} = 484$ and 674~ms). Care was taken to unambiguously determine $n$ for each ion. The resolved ground state and isomer in $^{136}$I are presented in Fig.~\ref{fig:136I}. More details about the technique can be found in Refs. \cite{Nesterenko2018,Nesterenko2021}. 

For cases where it was statistically feasible, the count-rate class analysis was performed \cite{Kellerbauer2003,Roux2013,Nesterenko2021} to account for ion-ion interactions in the precision trap. In case of the TOF-ICR measurements, the systematic uncertainties due to magnetic field fluctuations ${\delta B/B = 8.18(19) \times 10^{-12} \mathrm{~min}^{-1} \times \delta t}$ \cite{Canete2016}, with $\delta t$ being the time between measurements, and a mass-dependent uncertainty ${\delta r/r = 2.2(6) \times 10^{-10} / \textnormal{u} \times (M_{ref} - M)}$ \cite{Canete2019} were taken into account. In case of the PI-ICR measurements, the systematic uncertainties due to the phase advancement of different mass ions in $\phi_-$ measurement, the angle error, and the magnetic field fluctuations ${\delta B/B = 2.01(25) \times 10^{-12}\mathrm{~min}^{-1}\times \delta t}$ \cite{Nesterenko2021} were included. In addition, for $^{136}$I$^{m}$ and $^{140}$I measured against the $^{133}$Cs$^+$ reference ion, a mass-dependent uncertainty of ${\delta r/r = -2.35(81) \times 10^{-10} / \textnormal{u} \times (M_{ref} - M)}$ and a residual systematic uncertainty of ${\delta r/r=9\times 10^{-9}}$ were also added \cite{Nesterenko2021}.

\section{Results and discussion}

\begin{table*}
\caption{\label{tab:results}The frequency ratios $r = \nu_{c,ref}/\nu_{c}$ and corresponding mass-excess values $\Delta$ determined in this work. Reference ions (Ref.), the excitation times for the TOF-ICR technique $T_{trap}$ (accumulation times for the PI-ICR technique, marked with $^{t}$), the AME20 \cite{AME2020} values ($\Delta_{lit.}$) and differences between this work and AME20 (${\mathrm{Diff.} = \Delta-\Delta_{lit.}}$) are also added. Half-lives ($T_{1/2}$) and spin-parity assignments ($J^{\pi}$) are taken from NUBASE20 \cite{NUBASE20} except for $^{140}$I for which they are taken from Ref. \cite{Yagi2022}. The $^{140x}$I nuclide was assigned to a mixture of three long-lived states, see text for details. \# indicates spin-parity assignments based on systematics.}
\begin{ruledtabular}
\begin{tabular}{lllllllll}
Nuclide & $T_{1/2}$ & $J^{\pi}$ & Ref. & $T_{trap}$ (ms) & $r$ & $\Delta$ (keV) & $\Delta_{lit.}$ (keV) & Diff. (keV)  \\\hline
$^{137}$Sb & 497(21) ms & $7/2^+$\# & $^{136}$Xe & 25-150-25 & 1.007 566 669(50) & $-60011.3(63)$ & $-60060(50)$ & $49(50)$ \\
$^{136}$I & 83.4(4) s & ($1^-$) & $^{136}$I$^m$ & 300$^{t}$ & 0.999 998 301(34) & $-79550.5(44)$ & $-79545(14)$ & $-6(15)$\\
$^{136}$I$^m$ & 46.6(11) s & ($6^-$)&  $^{133}$Cs &  300$^{t}$ & 1.022 643 126(25) & $-79331.2(31)$ & & \\
& & &  $^{136}$Xe & 25-750-25 & 1.000 056 0335(48) & $-79335.54(61)$ & & \\
\multicolumn{6}{r}{weighted mean:} & $-79335.38(82)$ & $-79339(5)$ & $3.6(51)$\\
$^{137}$I & 24.13(12) s & $7/2^+$\# & $^{133}$Cs & 25-350-25, 25-750-25 & 1.030 191 2863(39) & $-76362.07(48)$ & $-76356(8)$ & $-6.1(80)$\\
$^{138}$I & 6.26(3) s & ($1^-$) & $^{133}$Cs & 25-350-25 & 1.037 750 9941(66)& $-71963.34(82)$ & & \\
& & & $^{130}$Xe & 25-350-25 & 1.061 732 5002(53) & $-71963.60(64)$ & & \\ 
\multicolumn{6}{r}{weighted mean:} & $-71963.50(50)$ & $-71980(6)$ & 17(6) \\ 
$^{139}$I & 2.280(11) s & $7/2^+$\# & $^{133}$Cs & 25-350-25& 1.045 303 392(18) & $-68469.6(22)$ & $-68471(4)$ & $1.4(46)$\\
$^{140}$I$^x$ & 0.38(2) s\footnotemark[1] & $(2^-,3)$\footnotemark[1] & $^{133}$Cs & 25-350-25 & 1.052 867 0435(51) & & &  \\
 &  &   & $^{133}$Cs & 484$^t$, 674$^t$ & 1.052 867 028(12) & &  &  \\
\multicolumn{5}{r}{weighted mean:}  & 1.052 867 0411(56) & $-63582.90(69)$ & $-63606(12)$ & 23(12) \\
$^{141}$I & 420(7) ms & $7/2^+$\# & $^{133}$Cs & 25-150-25 & 1.060 420 876(36) & $-59911.2(45)$ & $-59927(16)$ & $16(17)$\\
$^{142}$I & 235(11) ms & $2^-$\# & $^{133}$Cs & 100 & 1.067 985 99(48) & $-54843(59)$ & $-54803(5)$ & $-40(59)$ \\
\end{tabular}
\end{ruledtabular}
\footnotetext[1]{$T_{1/2}$ and $J^\pi$ are of the ground state, as reported in Ref. \cite{Yagi2022}. For the low-spin isomer $T_{1/2}$ = 0.91(5) s and $J^\pi = (0^-,1)$, while for the high-spin isomer $T_{1/2}$ = 0.47(4) s and $J^\pi = (4^-,5)$ \cite{Yagi2022}.}
\end{table*}

The summary of measured frequency ratios, deduced mass excess values and a comparison with the Atomic Mass Evaluation 2020 (AME20) values \cite{AME2020} are presented in Table~\ref{tab:results}. With the exception of $^{138}$I and $^{140}$I, the results from this work are in agreement with AME20 which is almost exclusively based on the results reported from CPT \cite{VanSchelt2012,VanSchelt2013,Orford2020}. The precision was improved for all the reported species except $^{142}$I \cite{Orford2020}. 

In the case of $^{137}$Sb, there are two contradictory mass measurements in the literature. The first one is reported by CPT, ${\Delta_{lit.} = -60061(52)}$~keV \cite{VanSchelt2013}, while the second one is reported by ESR, measured using the Isochronous Mass Spectrometry, ${\Delta_{lit.} = -60610(173)}$~keV \cite{Knobel2016}. The latter was rejected by the AME20 evaluators \cite{Huang2021}. The mass-excess value from this work, ${\Delta = -60011.3(63)}$~keV, agrees with the CPT result but it is eight times more precise. It also validates the decision of the AME20 evaluators to reject the ESR value. 

The mass of one state in $^{136}$I, assigned as an isomer in AME20 \cite{AME2020,Huang2021}, was measured at CPT \cite{VanSchelt2012}. In this work, $^{136}$I$^{m}$ was measured during Run I, with the TOF-ICR method, and Run II, with the PI-ICR method. The final mass-excess value, ${\Delta = -79335.38(82)}$~keV, agrees with the CPT result \cite{VanSchelt2012} while being five times more precise.

The mass of the ground state of $^{136}$I is currently known exclusively from $\beta$ end-point studies \cite{AME2020}. In this work it was measured during Run II against the isomer using the PI-ICR method with 300~ms accumulation time. The extracted excitation energy, ${E_x =  215.1(43)}$~keV agrees with the NUBASE20 value (${E_{x,lit.} =  206(15)}$~keV \cite{NUBASE20}) but it is more than three times more precise. It also leads to the ground-state mass-excess value of ${\Delta = -79550.5(43)}$~keV which agrees with the AME20 evaluation (${\Delta_{lit.} = -79545(14)}$~keV \cite{AME2020}) but is three times more precise.

The mass excess of $^{138}$I was measured against $^{130}$Xe and $^{133}$Cs during Runs I and II, respectively. Both measurements used the TOF-ICR method with an identical 25-350-25~ms (on-off-on) pattern and the results agree with each other. The final mass-excess value, $\Delta = -71963.50(50)$~keV, is 17(6)~keV (2.8 standard deviations) off from the CPT value ($\Delta_{lit.} = -71980(6)$~keV \cite{VanSchelt2012}) and it is 12 times more precise. The most likely explanation for this discrepancy is a presence of an unidentified isobaric contamination in one of the Penning traps. A similar issue was observed for $^{140}$Te: the old CPT value, reported in Ref. \cite{VanSchelt2013}, was three standard deviations off from the JYFLTRAP value \cite{Hakala2012}, while in the new CPT measurement, which used the PI-ICR method, a contaminant was discovered and the new value is consistent with the JYFLTRAP result \cite{Orford2018}.

\begin{figure}
\centering
\includegraphics[width=\columnwidth]{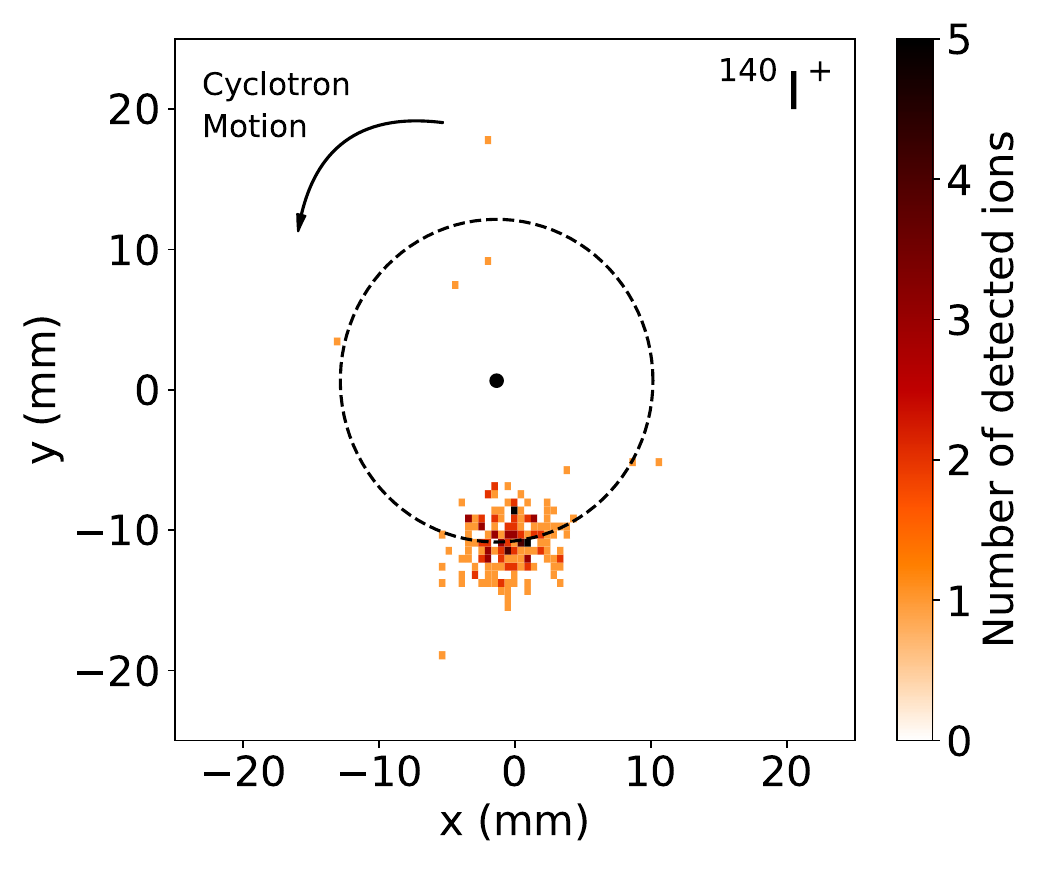}
\caption{\label{fig:140I}Projection of the cyclotron motion of $^{140}$I ions onto the position-sensitive detector obtained with the PI-ICR technique using a phase accumulation time $t_{acc} = 674$~ms. The average excitation radius is indicated with the dashed circle and the position of the center spot with the $\bullet$ symbol. Only one spot is present indicating that the isomers, if present, were not resolved.}
\end{figure}

The mass of $^{140}$I is known from the CPT measurement performed with the TOF-ICR technique \cite{VanSchelt2013} where only one state was observed. However, a more recent decay study performed at RIBF, RIKEN indicated a presence of three long-lived states in this nucleus, a ground state [$T_{1/2} = 0.38(2)$~s, $J^{\pi} = (2^-,3)$], a low-spin isomer [$0.91(5)$~s, $(0^-,1)$] and a high-spin isomer [$0.47(4)$~s, $(4^-,5)$] \cite{Yagi2022}. Since these states were not observed during the mass-measurement experiment, it might suggest that they were too close in energy to be resolved.

In this work the mass of $^{140}$I was measured in Run~II, using the TOF-ICR technique, and in Run III, using the PI-ICR technique, with $^{133}$Cs$^{+}$ ions used as a reference. In Run III, two different accumulation times were used, 484 and 674~ms. In both runs only one state was observed, see Fig. \ref{fig:140I}. The weighted average of the frequency ratio $r$ leads to the mass-excess value of ${\Delta = -63582.90(69)}$~keV which is 23(12)~keV less bound than the CPT result (${\Delta_{lit.} = -63606(12)}$~keV \cite{VanSchelt2013}). This discrepancy might be related to different production mechanisms and, consequently, a different isomeric ratio in the beam. Considering the 674~ms accumulation time used for the PI-ICR measurement in this work, the excitation energy of both isomers is estimated to be below 30 keV.

We note that with the spin-parity assignments for the three long-lived states, there must be at least one potential $M2$ transition between them\footnote{In principle, lower multipolarities ($E1$, $M1$, $E2$) might be also possible, however, this would lead to half-lives orders of magnitude shorter than hundreds of milliseconds reported in Ref. \cite{Yagi2022}.}. Considering that the reported half-lives are of the order of hundreds of milliseconds, the energy of this $M2$ transition can be expected to be of the order of a few tens of keV or lower \cite{Garg2023} which is below the resolving power reached in this work.

Since for the iodine isotopes the determined mass values are either in agreement with, or very close to, the literature values (see Table~\ref{tab:results}), the effect of the new masses on the calculated neutron-capture rates for the r process is almost negligible as compared to AME20 and is therefore not discussed further. The biggest difference is observed for $^{137}$Sb where the updated mass value leads also to a more precise neutron-capture $Q$-value, $Q_{n,\gamma}=3576(9)$~keV, as compared to the AME20 value, $Q_{n,\gamma}=3620(50)$~keV \cite{AME2020}. Our $Q$-value rules out the value based on the $^{137}$Sb mass from the ESR measurement, $Q_{n,\gamma}=4170(170)$~keV. 

\begin{figure}
\centering
\includegraphics[width=\columnwidth]{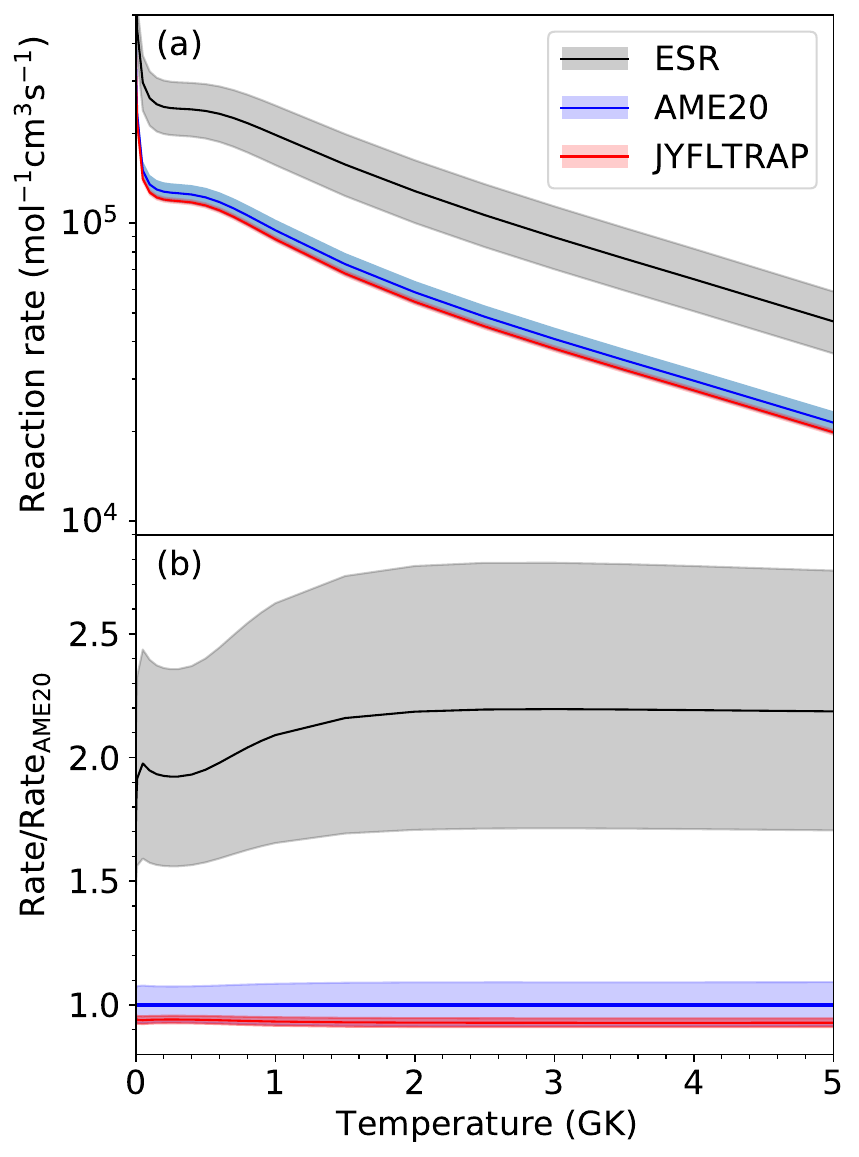}
\caption{\label{fig:137Sb}(a) Astrophysical reaction rates calculated with the Talys 1.96 code \cite{Koning2023} for the reaction $^{136}$Sb$(n,\gamma)^{137}$Sb using the mass value for $^{137}$Sb from this work (JYFLTRAP), AME20 \cite{AME2020} and ESR \cite{Knobel2016}. The AME20 value is based on the CPT result \cite{VanSchelt2013} and the mass of $^{136}$Sb was adopted from AME20 for all three cases. (b) Reaction rate ratios with respect to the AME20 rate.}
\end{figure}

The impact on the neutron-capture $^{136}$Sb$(n,\gamma)^{137}$Sb reaction rate was studied using the Talys 1.96 code \cite{Koning2023}. A comparison between this work (JYFLTRAP), AME20 \cite{AME2020}, and ESR \cite{Knobel2016} is presented in Fig.~\ref{fig:137Sb} for temperatures $T\leq5$~GK relevant for neutron captures in the r process (see e.g. Refs.~\cite{Cowan2021,Arnould2020}). At higher temperatures, photodisintegrations dominate and abundances are determined by the nuclear statistical equilibrium. The reaction rate obtained with the $^{137}$Sb mass value from this work and AME20 agree with each other. However, for example at a characteristic r-process temperature of $T=1$~GK \cite{Arnould2020}, the former is around 7\% lower than the latter and, in addition, it is around five times more precise. The reaction rate calculated with the ESR result significantly deviates from both, this work and AME20, and it is about twice larger. Thus, our measurement rules out such a neutron-capture reaction rate.

\section{Summary}

We have reported on the Penning-trap measurements of $^{137}$Sb and $^{136-142}$I performed with the JYFLTRAP double Penning trap mass spectrometer. The results reported in this work improved the precision compared to the literature values for all the studied isotopes with the exception of $^{142}$I, which was recently reported from CPT \cite{Orford2020}. In general, the results are in good agreement with the CPT measurements \cite{VanSchelt2012,VanSchelt2013,Orford2020}.

For $^{137}$Sb, our result agrees with the previous Penning-trap mass measurement from CPT \cite{VanSchelt2013} but is eight times more precise. This allows us to firmly rule out the mass value obtained for $^{137}$Sb at the ESR \cite{Knobel2016}. The obtained neutron-capture rate for $^{136}$Sb$(n,\gamma)^{137}$Sb is in agreement with the rate calculated with AME20 \cite{AME2020} but is around five times more precise at $T=1$~GK. 
For $^{136}$I, the ground state mass and the isomer excitation energy were measured directly for the first time. The latter is in agreement with NUBASE20 \cite{NUBASE20} but is more than three times more precise. Our result also confirms the AME20 assignment \cite{Huang2021} that the CPT mass value for $^{136}$I \cite{VanSchelt2012} belongs to the isomeric state. 

The mass values for $^{138,140}$I reported in this work differ from the CPT measurements \cite{VanSchelt2012,VanSchelt2013} by 17(6)~keV and 23(12)~keV, respectively. A plausible explanation for the discrepancy between the $^{138}$I mass values could be an unresolved contaminant ion shifting the measured value. At the same time, in the case of $^{140}$I a discrepancy between the two Penning-trap values might be related to different isomeric yield ratios linked to different beam production methods. Recently, three long-lived states in $^{140}$I were observed in the $\beta$-decay study \cite{Yagi2022} and while they were not resolved in this work, nor at CPT, a 30~keV excitation energy limit was deduced in this work. 

We have improved the knowledge of the nuclear masses north-east of doubly magic $^{132}$Sn. Although the changes to the literature values are less than 50~keV, the measurements improve the precision for most of the studied cases and help to make more accurate r-process calculations. The observed differences for $^{138,140}$I show that it is important to cross-check the results from different facilities to obtain not only precise but more accurate results. \\

\begin{acknowledgments}

This project has received funding from the European Union’s Horizon 2020 Research and Innovation Programme under Grant Agreements No. 771036 (ERC CoG MAIDEN) and No. 861198–LISA–H2020-MSCA-ITN-2019, from the European Union’s Horizon Europe Research and Innovation Programme under Grant Agreement No. 101057511 (EURO-LABS), and from the Academy of Finland projects No. 295207, No. 306980, No. 327629, No. 354589 and No. 354968. J.R. acknowledges financial support from the Vilho, Yrj\"o and Kalle V\"ais\"al\"a Foundation. 

\end{acknowledgments}

\bibliographystyle{apsrev}
\bibliography{sn132bib}

\end{document}